\documentclass[aps,prd,superscriptaddress,twocolumn]{revtex4-1}
\usepackage{amssymb}
\usepackage{amsmath}
\usepackage{wallpaper}
\usepackage{bm}
\usepackage{multirow}
\usepackage[colorlinks,linkcolor=blue, urlcolor=blue, anchorcolor=blue, citecolor=blue]{hyperref}

\begin{document}

\title{Finite nuclei in an extended Nambu-Jona-Lasinio model}

\author{Cheng-Jun Xia}
\email{cjxia@yzu.edu.cn}
\affiliation{Center for Gravitation and Cosmology, College of Physical Science and Technology, Yangzhou University, Yangzhou 225009, China}

\author{Yu-Ting Rong}
\email{rongyuting@gxnu.edu.cn}
\affiliation{Guangxi Key Laboratory of Nuclear Physics and Technology, Guangxi Normal University, Guilin, 541004, China}

\author{Ting-Ting Sun}
\email{ttsunphy@zzu.edu.cn}
\affiliation{School of Physics, Zhengzhou University, Zhengzhou 450001, China}

\date{\today}

\begin{abstract}
We propose a new theoretical framework to investigate the properties of finite nuclei based on an extended Nambu-Jona-Lasinio (eNJL) model, where the Dirac sea, the spontaneous chiral symmetry breaking, and the quark degrees of freedom are considered by extending the SU(3) NJL model and treating baryons as clusters of quarks. The eNJL model can then be readily adopted to examine the matter states ranging from baryonic matter to quark matter in a unified manner. In this work, by assuming spherically symmetric finite nuclei and neglecting the center-of-mass or rotational corrections, we systematically investigate the properties of finite nuclei based on the eNJL model with additional pairing correlations. It is found that our model generally reproduces the binding energies of the 2495 nuclei ($A>2$) from the 2016 Atomic Mass Evaluation (AME2016) with the root-mean-square deviations $5.38$ MeV. The deviations are mainly attributed to the too large shell gaps at magic numbers $N(Z) =28$, 50, and 82 as well as the spurious shell closures at $N(Z)=34$, 58, and 92. Meanwhile, the obtained charge radii of 906 nuclei are systematically smaller than the experimental values with root-mean-square deviations $0.127$ fm. In our future study, we expect to reduce the uncertainties of our predictions by carefully calibrating the density dependence of coupling constants and considering deformations with microscopic collective corrections from the nucleons in the Fermi sea and quarks in the Dirac sea.
\end{abstract}

\maketitle

\section{\label{sec:intro}Introduction}

By treating nucleons as the basic building blocks, it was shown that the properties of finite nuclei and nuclear matter below and around the nuclear saturation density $n_{0}\approx 0.16$ fm${}^{-3}$ can be well understood, e.g., using relativistic density functional theories~\cite{Meng2016_RDFNS}, nonrelativistic density functional theories~\cite{Dutra2012_PRC85_035201}, or microscopic many-body theories~\cite{Hebeler2021_PR890-1}. Being the fundamental constituents of nucleons, the quark degrees of freedom seems irrelevant and nucleons can be treated as point particles since exciting a nucleon takes $\sim$300 MeV while exciting a nucleus takes $\sim$10 MeV. Nevertheless, in relativistic density functional theories~\cite{Meng2016_RDFNS}, there exist strong attractive scalar potential ($\sim$440 MeV) and repulsive vector potential ($\sim$360 MeV) to accommodate the spin-orbit splitting of single particle energies in finite nuclei. In microscopic many-body theories~\cite{Hebeler2021_PR890-1}, the many-body interactions were shown to be important to achieve the saturation properties of nuclear matter, which effectively account for the internal excitations of nucleons. In fact, the structures of nucleons in nuclear medium deviate from those of free nucleons, e.g., the European Muon Collaboration (EMC) effect~\cite{Aubert1983_PLB123-275}. It is thus favorable to consider directly the quark degrees of freedom in nuclear matter.

To account for the contributions of quarks in nucleons, the quark-meson-coupling (QMC) model was then proposed considering the effects of the three valence quarks confined in bags so that the EMC effect can be understood~\cite{Guichon2018_PPNP100-262}. Nevertheless, at larger densities, the boundaries between baryons become less distinctive as the number of exchanged quarks increases~\cite{Oka1980_PLB90-41, Chen2007_PRC76-014001, Fukushima2016_ApJ817-180, McLerran2009_NPA830-709c, Park2021_PRD104-094024, Park2022_PRD105-114034}. It is thus insufficient to treat nucleons as independent particles and a unified description based on quark degrees of freedom is favorable. Extensive efforts were then devoted to this field in the past decades, e.g., the string-flip model~\cite{Horowitz1991_PRC44-2753}, color molecular dynamics~\cite{Maruyama2000_PRC61-062201, Yasutake2024_PRD109-043056}, quark-diquark approach in the framework of Nambu-Jona-Lasinio (NJL) model~\cite{Bentz2001_NPA696-138}, and the hadronic SU(3) nonlinear $\sigma$ model including quarks~\cite{Dexheimer2010_PRC81-045201}.
A direct modeling of quarkyonic matter was recently proposed~\cite{McLerran2019_PRL122-122701, Margueron2021_PRC104-055803}, while a complete field model was later developed by synthesizing the Walecka model and quark-meson model~\cite{Cao2020_JHEP10-168, Cao2022_PRD105-114020}. By combining relativistic mean field (RMF) models and equivparticle models with density-dependent quark masses, the quarkyonic matter and quarkyonic transition were investigated as well~\cite{Xia2018_JPSCP20-011010, Xia2023_PRD108-054013}.

In our previous study, by considering baryons as cluster of quarks we have extended the NJL (eNJL) model to describe baryonic matter, quark matter, and their transitions in a unified manner, where various first-order phase transitions and the microscopic structures of thier mixed phases were investigated~\cite{Xia2024_PRD110-014022, Xia2024}. The Dirac sea, the spontaneous chiral symmetry breaking, and the quark degrees of freedom can then be treated in a unified manner in the framework of the eNJL model. In this work, based on the eNJL model, we further investigate the properties of finite nuclei in mean field approximation (MFA) and solve the Dirac equations of nucleons assuming nuclei are spherically symmetric. It is found that our model generally reproduces the binding energies of the 2495 nuclei ($A>2$) from the 2016 Atomic Mass Evaluation (AME2016)~\cite{Audi2017_CPC41-030001, Huang2017_CPC41-030002, Wang2017_CPC41-030003}, while the obtained charge radii of 906 nuclei are systematically smaller than the experimental values~\cite{Angeli2013_ADNDT99-69}.

The paper is organized as follows. In Sec.~\ref{sec:the} we present our theoretical framework, including the Lagrangian density of the eNJL model and the formalism of fixing the properties of spherically symmetric finite nuclei in MFA. The obtained properties of finite nuclei are presented in Sec.~\ref{sec:res}.  We draw our conclusion in Sec.~\ref{sec:con}.

\section{\label{sec:the}Theoretical framework}
\subsection{\label{sec:the_Lagrangian} Lagrangian density}
The Lagrangian density of the eNJL model~\cite{Xia2024_PRD110-014022, Xia2024} reads
\begin{eqnarray}
\mathcal{L} &=& \sum_{i} \bar{\psi}_i \left(  i \gamma_\mu D^\mu_i - M_i \right)\psi_{i} +4K
 \bar{n}_{u}^{s}\bar{n}_{d}^{s}\bar{n}_{s}^{s} \label{eq:Lagrangian}\\
 &&\mbox{} - \frac{1}{4} \omega_{\mu\nu}\omega^{\mu\nu}  + \frac{1}{2}m_\omega^2 \omega^2 - \frac{1}{4} \vec{\rho}_{\mu\nu}\cdot\vec{\rho}^{\mu\nu}  + \frac{1}{2}m_\rho^2 \rho^2   \nonumber \\
 &&\mbox{}  + \frac{1}{2}\sum_{q=u,d,s} \left[
\partial_\mu \sigma_q \partial^\mu \sigma_q - m_\sigma^2 \sigma_q^2 \right] - \frac{1}{4} A_{\mu\nu}A^{\mu\nu},  \nonumber
\end{eqnarray}
where $\psi_{i}$ is the Dirac spinor for fermion $i$ ($=n,p,u,d,s$). Compared with the SU(3) NJL model~\cite{Rehberg1996_PRC53-410}, the quark condensations $\langle \bar{\psi}_{q} \psi_{q} \rangle$ (${q=u,d,s}$) are replaced by  $\sigma_q$ fields, then the four-point interaction becomes nonlocal and the interaction range is modulated by $m_\sigma$. The six-point (`t Hooft) interaction $4K  \bar{n}_{u}^{s}\bar{n}_{d}^{s}\bar{n}_{s}^{s}$ remains local, where the effective quark scalar density is fixed by
\begin{equation}
  \bar{n}_q^s= n_{q}^{s}+\alpha_S \sum_{i=p,n} N^q_i  n_{i}^{s}. \label{eq:bnqs}
\end{equation}
Here $N^q_i$ stands for the number of valence quarks $q$ in particle $i$ and $n_{i}^{s}=\langle \bar{\psi}_{i} \psi_{i} \rangle$. In addition, vector mesons and photons are considered in the eNJL model as well, where the field tensors $\omega_{\mu\nu}$, $\vec{\rho}_{\mu\nu}$, and $A_{\mu\nu}$ are obtained with
\begin{eqnarray}
\omega_{\mu\nu} &=& \partial_\mu \omega_\nu - \partial_\nu \omega_\mu,  \\
\vec{\rho}_{\mu\nu} &=& \partial_\mu \vec{\rho}_\nu - \partial_\nu \vec{\rho}_\mu,  \\
A_{\mu\nu} &=& \partial_\mu A_\nu - \partial_\nu A_\mu.
\end{eqnarray}
The covariant derivative is given by
\begin{equation}
i D^\mu_i = i \partial^\mu -  g_{\omega} \sum_{q=u,d,s} N^q_i \omega^\mu  - g_{\rho} \vec{\tau}_i\cdot\vec{\rho}^\mu - e q_i A^\mu,
\end{equation}
where $\vec{\tau}_i$ is the isospin and $q_i$ the charge number of fermion $i$ with $q_{n}=0$, $q_{p}=e$, $q_{u}=2e/3$, and $q_{d}=q_{s}=-e/3$. In eNJL model, baryons are considered as clusters with the effective masses fixed by
\begin{equation}
 M_i = \sum_{q=u,d,s} N^q_i \left[m_{q0} + \alpha_S(M_{q}-m_{q0})\right] \label{eq:Bmass}
\end{equation}
with the quark mass
\begin{equation}
 M_{q} = m_{q0} - g_{\sigma} \sigma_q + 2 K \frac{\bar{n}_{u}^{s}\bar{n}_{d}^{s}\bar{n}_{s}^{s}}{\bar{n}_q^{s}}. \label{eq:qmass}
\end{equation}
Here $m_{u0}= m_{d0} = 5.5$ MeV and $m_{s0}=  140.7$ MeV represent the current masses of quarks~\cite{Rehberg1996_PRC53-410}. To reproduce the saturation properties of nuclear matter, the couplings constants are density-dependent and take the following forms~\cite{Xia2024_PRD110-014022}, i.e.,
\begin{eqnarray}
  \alpha_S &=& a_S \exp(-n_\mathrm{b}/n_S)+b_S,  \label{eq:alphaS} \\
g_\omega^2 &=& 4G_S m_\omega^2 [a_V\exp(-n_\mathrm{b}/n_V) + b_V], \label{eq:alphaV}\\
g_\rho^2   &=& 4G_S m_\rho^2 [a_{TV}\exp(-n_\mathrm{b}/n_{TV}) + b_{TV}], \label{eq:alphaTV}
\end{eqnarray}
where the corresponding parameters of the eNJL model are indicated in Table~\ref{table:DDparam}. Note that $G_S$ and $K$ are the coupling strengths for the four-point and six-point interactions in the SU(3) NJL model, where we have adopted the RKH parameter set $G_S = g_{\sigma}^2/4m_{\sigma}^2= 1.835/\lambda^2$ and $K = 12.36/\lambda^5$ with $\lambda = 602.3$ MeV being the 3-momentum cut-off to regularize the vacuum part of quarks~\cite{Rehberg1996_PRC53-410}.

\begin{table}
  \centering
  \caption{\label{table:DDparam} Parameters for the eNJL model~\cite{Xia2024_PRD110-014022}, which reproduce the nuclear saturation properties and finite nuclei properties adopting the RKH parameter set~\cite{Rehberg1996_PRC53-410}.}
  \begin{tabular}{l|l|l}
    \hline \hline
  $a_S=0.4413715$     & $n_S=0.16\ \mathrm{fm}^{-3}$    & $b_S=0.4076285$ \\
  $a_V=3.566049$     & $n_V=0.214\ \mathrm{fm}^{-3}$   & $b_V=1.062771$ \\
  $a_{TV}=0.5014459$  & $n_{TV}=0.1\ \mathrm{fm}^{-3}$  & $b_{TV}=0.0117601$ \\     \hline
  $m_{\sigma}=525$ MeV  & $m_{\omega}=2500$ MeV  & $m_{\rho}=769$ MeV  \\        \hline
  \end{tabular}
\end{table}

\subsection{\label{sec:the_Dirac} Finite nuclei}
In MFA, the boson fields $\sigma_i$, $\omega_\mu$, $\vec{\rho}_\mu$, and $A_\mu$ take mean values and are left with only the time components due to time-reversal symmetry, while charge conservation guarantees that only the $3$rd component (${\rho}_{\mu,3}$ and $\tau_{i,3}$) in the isospin space remains. We then define $\omega\equiv \omega_0$, $\rho\equiv \rho_{0,3}$, and $\tau_i\equiv \tau_{i,3}$ for simplicity. To fix the properties of finite nuclei in MFA, we assume they are spherically symmetric so that the Dirac spinors can be expanded as
\begin{equation}
 \psi_{n\kappa m}({\bm r}) =\frac{1}{r}
 \left(\begin{array}{c}
   iG_{n\kappa}(r) \\
    F_{n\kappa}(r) {\bm\sigma}\cdot{\hat{\bm r}} \\
 \end{array}\right) Y_{jm}^l(\theta,\phi)\:,
\label{EQ:RWF}
\end{equation}
with $G_{n\kappa}(r)/r$ and $F_{n\kappa}(r)/r$ being the radial wave functions for the upper and lower components, while $Y_{jm}^l(\theta,\phi)$
is the spinor spherical harmonics. The quantum number $\kappa$ is connected to the angular momenta $(l,j)$ via $\kappa=(-1)^{j+l+1/2}(j+1/2)$. The Dirac equation for the radial wave functions of nucleons ($i=n,p$) is then
\begin{equation}
 \left(\begin{array}{cc}
  V_i+M_i                             & {\displaystyle -\frac{\mbox{d}}{\mbox{d}r}+\frac{\kappa}{r}}\\
  {\displaystyle \frac{\mbox{d}}{\mbox{d}r}+\frac{\kappa}{r}} & V_i-M_i                       \\
 \end{array}\right)
 \left(\begin{array}{c}
  G_{n\kappa} \\
  F_{n\kappa} \\
 \end{array}\right)
 = \varepsilon_{n\kappa}
 \left(\begin{array}{c}
  G_{n\kappa} \\
  F_{n\kappa} \\
 \end{array}\right) \:,
\label{EQ:RDirac}
\end{equation}
with the single particle energy $\varepsilon_{n\kappa}$ and the mean field vector potential
\begin{equation}
  V_i = \sum_{q=u,d,s} N^q_i  g_{\omega} \omega +  g_{\rho} \tau_{i}  \rho  +\Sigma^\mathrm{R}+e q_i  A.  \label{EQ:potential} 
\end{equation}
Note that the mean field scalar potential $S_i$ has been considered in the effective mass in Eq.~(\ref{eq:Bmass}), where $M_i$ decreases in nuclear medium and $S_i=M_i-M_i^0$ with $M_i^0$ being the nucleon mass in vacuum. The ``rearrangement'' term in Eq.~(\ref{EQ:potential}) arises due to the density dependence of coupling constants and is given by
\begin{eqnarray}
 \Sigma^\mathrm{R}&=& \sum_{i=n,p} \left( \omega n_{i}
 \sum_{q=u,d,s} N^q_i \frac{\mbox{d} g_{\omega}}{\mbox{d} n_\mathrm{b}} +  \rho \tau_{i} n_{i} \frac{\mbox{d} g_{\rho}}{\mbox{d} n_\mathrm{b}}\right)
  \nonumber \\
 && +\sum_{i=n,p} \left[ \frac{\mbox{d}  \alpha_S}{\mbox{d} n_\mathrm{b}}\sum_{q=u,d,s} N^q_i(M_{q}-m_{q0}) \right] n_{i}^s. \label{eq:Sigma_b}
\end{eqnarray}
The boson fields are fixed by solving the following equations, i.e.,
\begin{eqnarray}
(-\nabla^2 + m_\sigma^2) \sigma_q &=&  g_{\sigma} \bar{n}_q^s, \label{eq:KG_sigma} \\
(-\nabla^2 + m_\omega^2) \omega &=&  \sum_{i=n,p} \sum_{q=u,d,s} N^q_i  g_{\omega} n_i, \label{eq:KG_omega}\\
(-\nabla^2 + m_\rho^2) \rho     &=& g_{\rho} \sum_{i=n,p}   \tau_i n_i, \label{eq:KG_rho}\\
                   -\nabla^2 A  &=& e\sum_{i=n,p}q_i n_i. \label{eq:KG_photon}
\end{eqnarray}
Here $\bar{n}_q^s$ is fixed by Eq.~(\ref{eq:bnqs}). The source currents $n_{i} = \langle \bar{\psi}_{i} \gamma^{0} \psi_{i} \rangle$ and $n_{i}^{s} = \langle \bar{\psi}_{i} \psi_{i} \rangle$ of nucleon $i$ can be fixed according to the radial wave functions, i.e.,
\begin{subequations}
\begin{eqnarray}
 n_{i}^{s}(r) &=& \frac{1}{4\pi r^2}\sum_{k=1}^{N_i}
 \left[|G_{k i}(r)|^2-|F_{k i}(r)|^2\right] \:,
\\
 n_{i}(r) &=& \frac{1}{4\pi r^2}\sum_{k=1}^{N_i}
 \left[|G_{k i}(r)|^2+|F_{k i}(r)|^2\right] \:.
\end{eqnarray}%
\label{EQ:Density}%
\end{subequations}
Meanwhile, the contribution of quarks in the Dirac sea needs to be considered. In this work, instead of solving the wave functions of quarks, for simplicity, we adopt Thomas-Fermi approximation (TFA) for quarks in the Dirac sea and the scalar density for quark $q$ is then determined by
\begin{equation}
n_{q}^{s}=\langle \bar{\psi}_{q} \psi_{q} \rangle = - \frac{3 M_{q}^{3}}{2\pi^{2}}\left[ y_{q}\sqrt{y_{q}^{2}+1}+\mathrm{arcsh}(y_{q})\right], \label{eq:ns}
\end{equation}
where $y_{q} \equiv \lambda/M_{q}$ with $\lambda = 602.3$ MeV~\cite{Rehberg1996_PRC53-410}. Note that nucleons can not exist in Dirac sea since quarks no longer form clusters due to Pauli blocking. The nucleon numbers can be calculated by integrating the baryon density $n_{i}(r)$ in coordinate space as
\begin{eqnarray}
Z=N_p &=&   \int 4\pi r^2 n_{p}(r) \mbox{d}r, \\
N =N_n &=& \int 4\pi r^2 n_{n}(r) \mbox{d}r,
\end{eqnarray}
where the total mass number is fixed with $A=Z+N$.

Finally, the total energy of the system in MFA is fixed by
\begin{equation}
E_\mathrm{MF}=\int \langle {\cal{T}}_{00} \rangle \mbox{d}^3 r, \label{eq:energy}
\end{equation}
with the energy momentum tensor
\begin{eqnarray}
\langle {\cal{T}}_{00} \rangle
&=&  \sum_{i=n,p}\sum_{k=1}^{N_i} \varepsilon_{ki} \langle \bar{\psi}_{ki} \gamma^{0} \psi_{ki} \rangle - \sum_{i=n,p}  V_i n_i -4K \bar{n}_{u}^{s}\bar{n}_{d}^{s}\bar{n}_{s}^{s} \nonumber \\
&&   + \frac{1}{2}\left[(\nabla \omega)^2 + m_\omega^2 \omega^2  + (\nabla \rho)^2 + m_\rho^2 \rho^2 +(\nabla A)^2\right]  \nonumber \\
&&  -\sum_{q=u,d,s} \frac {d_q{M_q}^4}{16\pi^{2}} \left[y_q(2y_q^2+1)\sqrt{y_q^2+1}-\mathrm{arcsh}(y_q) \right] \nonumber \\
&&      + \frac{1}{2}\sum_{q=u,d,s}\left[(\nabla \sigma_q)^2 +m_\sigma^2 \sigma_q^2\right] - \mathcal{E}_0. \label{eq:ener_dens}
\end{eqnarray}
Here a constant $\mathcal{E}_0$ is introduced to ensure $\langle {\cal{T}}_{00} \rangle = 0$ in the vacuum. Additionally, the pairing correlations need to be considered to account the odd-even staggering. Instead of solving the BCS equation or taking Bogoliubov transformation, for simplicity, in this work we adopt a phenomenological formula~\cite{Zubov2009_PPN40-847}, i.e.,
\begin{equation}
E_\mathrm{pair}=
\left\{
  \begin{array}{ll}
    -\delta/\sqrt{A}, & \text{even-even;} \\
    0, & \text{odd-even;} \\
    \delta/\sqrt{A}, & \text{odd-odd.}
  \end{array}
\right.
\label{EQ:pair}
\end{equation}
with $\delta = 6$ MeV. The total mass of a nucleus with given $Z$ and $A$ is then fixed by adding both $E_\mathrm{MF}$ and $E_\mathrm{pair}$, i.e.,
\begin{equation}
E_\mathrm{tot} = E_\mathrm{MF} + E_\mathrm{pair}.  \label{eq:etot}
\end{equation}
The binding energy of the nucleus is then determined by
\begin{equation}
B = M_N^0 A - E_\mathrm{tot}, \label{eq:bind}
\end{equation}
where $M_N^0=938.9$ MeV is the nucleon mass in vacuum obtained with Eq.~(\ref{eq:Bmass}). Note that the center-of-mass corrections $E_\mathrm{cm}$ was neglected in Eq.~(\ref{eq:etot}), which should be included in our future study~\cite{Bender2000_EPJA7-467}. Nevertheless, we expect a cancelation between the center-of-mass corrections from nucleons in the Fermi sea (negative) and quarks in the Dirac sea (positive), so that $E_\mathrm{cm}$ become relatively small.

Based on the density profiles, the charge radius $R$ can be fixed by~\cite{Long2004_PRC69-034319}
\begin{equation}
  R^2 = \frac{A-2}{A} R_p^2 + \frac{1}{A} R_M^2 + (0.862\ \mathrm{fm})^2 - (0.336\ \mathrm{fm})^2\frac{N}{Z},
\end{equation}
with
\begin{eqnarray}
R_p^2 &=& \int 4\pi r^4 n_p(r) \mbox{d}r/Z, \\
R_M^2 &=& \int 4\pi r^4 [n_p(r)+n_n(r)] \mbox{d}r/A.
\end{eqnarray}

\section{\label{sec:res}Results and discussions}
For fixed neutron $(N)$ and proton $(Z)$ numbers, we then solve the Dirac Eq.~(\ref{EQ:RDirac}), nucleon mass Eq.~(\ref{eq:Bmass}), quark mass Eq.~(\ref{eq:qmass}), mean field vector potentials Eq.~(\ref{EQ:potential}), boson field Eqs.~(\ref{eq:KG_sigma}-\ref{eq:KG_photon}), and densities Eq.~(\ref{EQ:Density}) in the eNJL model iteratively in coordinate space with a box size of $12.8~{\rm fm}$ and a grid distance of $0.1~{\rm fm}$. Once convergency is reached, the energy and binding energy can be fixed by Eqs.~(\ref{eq:etot}) and (\ref{eq:bind}).

\begin{figure}[!ht]
  \centering
  \includegraphics[width=\linewidth]{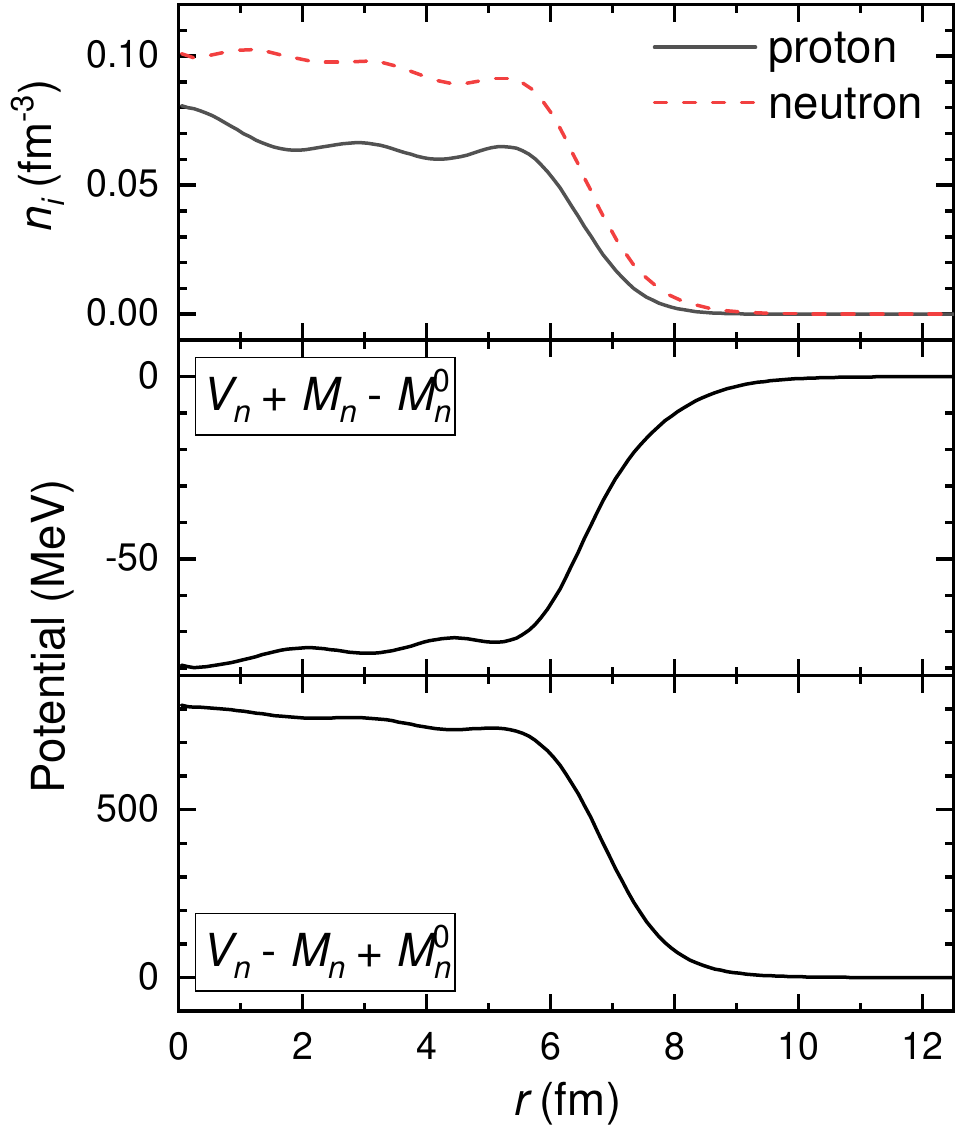}
  \caption{\label{Fig:Potential_Pb208}Density profiles and mean field potentials of $^{208}$Pb obtained with the eNJL model, where the parameters indicated in Table.~\ref{table:DDparam} are adopted.}
\end{figure}

As an example, in Fig.~\ref{Fig:Potential_Pb208} we present the obtained density profiles and mean field potentials of $^{208}$Pb, where the adopted parameters are indicated in Table.~\ref{table:DDparam}. It is found that the potential depths and density profiles are generally consistent with those obtained with RMF models, which would lead to correct description of finite nuclei. Note that we have used a rather large $\omega$ meson mass $m_{\omega}=2500$ MeV to prevent unrealistic density fluctuations caused by the six-point (`t Hooft) interaction, where $m_{\omega}$ should be viewed as an effective $\omega$ meson mass including the contributions of the `t Hooft term. Meanwhile, since the quantum fluctuations of nucleons have been considered, here $m_{\omega}$ can be smaller than those in Ref.~\cite{Xia2024} adopting TFA for nucleons.

\begin{figure}[!ht]
  \centering
  \includegraphics[width=\linewidth]{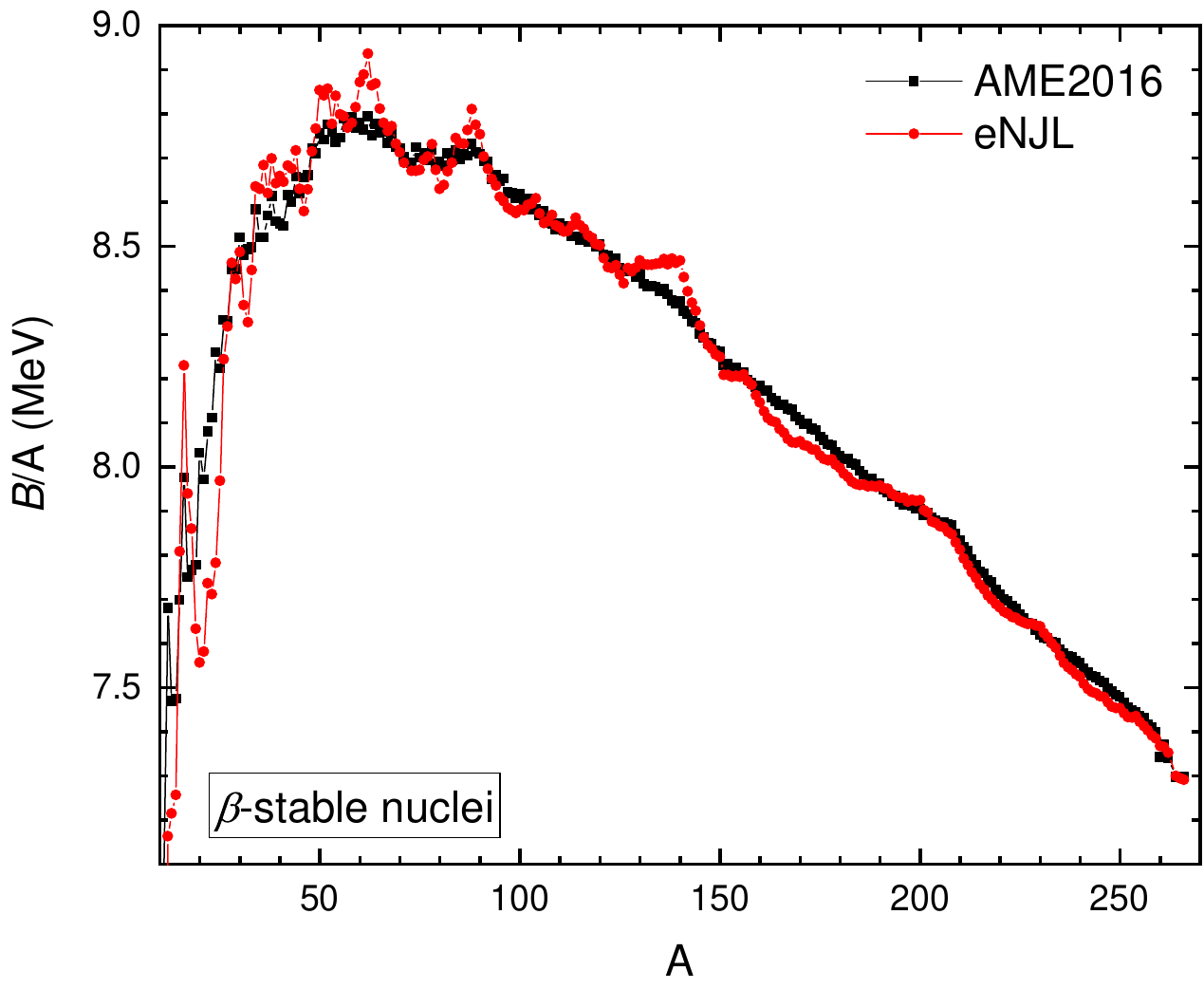}
  \caption{\label{Fig:B_Nucl_beta} Binding energy per nucleon of finite nuclei obtained with the eNJL model, which are compared with the experimental data from AME2016~\cite{Audi2017_CPC41-030001, Huang2017_CPC41-030002, Wang2017_CPC41-030003}.}
\end{figure}

Based on the parameter set indicated in Table.~\ref{table:DDparam}, we then carry out systematic investigations on the properties of finite nuclei in the framework of eNJL model. In Fig.~\ref{Fig:B_Nucl_beta} we present the binding energies per nucleon ($B/A$) of $\beta$-stable nuclei, which are compared with the experimental values from AME2016~\cite{Audi2017_CPC41-030001, Huang2017_CPC41-030002, Wang2017_CPC41-030003}. The obtained binding energies for $\beta$-stable nuclei are generally consistent with the experimental values with slight deviations at certain mass numbers, e.g., at $A=16$, 20, 32, 50, 62, 80, 88, 140, etc.

\begin{figure*}[!ht]
  \centering
  \includegraphics[width=0.7\linewidth]{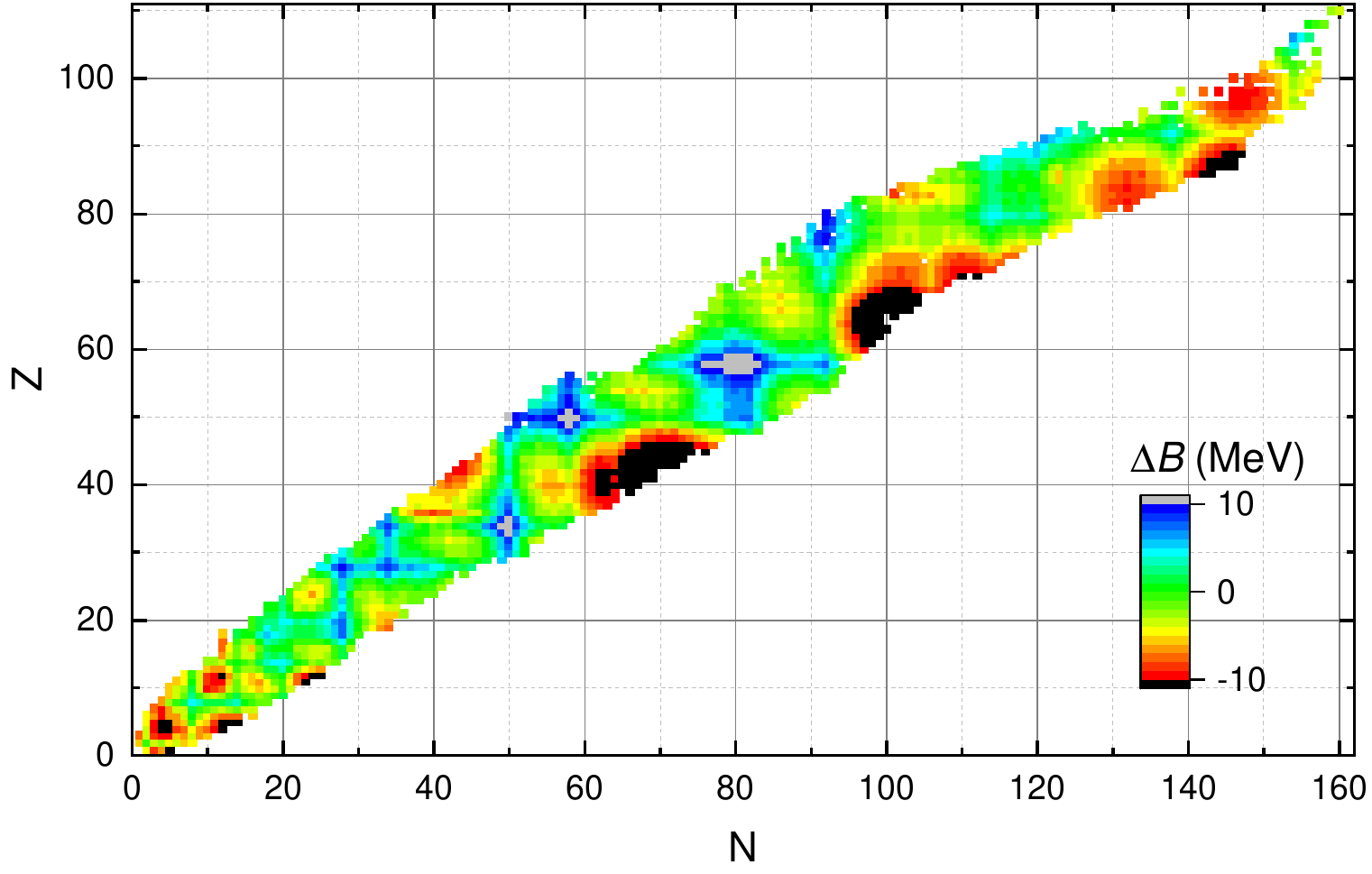}
  \caption{\label{Fig:DBall2D} The deviation of the theoretical binding energies ($\Delta B = B - B_\mathrm{exp}$) of the 2495 nuclei from the experimental values ($B_\mathrm{exp}$) in AME2016~\cite{Audi2017_CPC41-030001, Huang2017_CPC41-030002, Wang2017_CPC41-030003}.}
\end{figure*}

To show the the deviations more clearly, in Fig.~\ref{Fig:DBall2D} we present the differences ($\Delta B = B - B_\mathrm{exp}$) between the theoretical binding energies and the experimental values of the 2495 nuclei in AME2016~\cite{Audi2017_CPC41-030001, Huang2017_CPC41-030002, Wang2017_CPC41-030003}. It is found that our model generally reproduces the binding energies of the 2495 nuclei ($A>2$) from AME2016 with the root-mean-square deviations $\sqrt{\sum_i \Delta B_i^2/2495} = 5.38$ MeV. Evidently, as indicated in Fig.~\ref{Fig:DBall2D}, the nuclei around the magic numbers $N(Z)=28$, 50, and 82 are over bound, where the binding energy may exceed the experimental values by up to $\sim$10 MeV for double magic nuclei. Additionally, in our model we have observed spurious shell closures at $N(Z)=34$, 58, and 92, leading to over bound nuclei in these regions as well. Beside the over bound nuclei at magic numbers and spurious shell closures, the binding energy for nuclei at $Z\approx 4$, 12, 44, 65, 87 and $N\approx 4$, 12, 25, 44, 72, 101, 146 are too small compared with the experimental values. As will be illustrated later, those deviations are caused by too large shell closures with the nuclei being too compact.

\begin{figure}[!ht]
  \centering
  \includegraphics[width=0.8\linewidth]{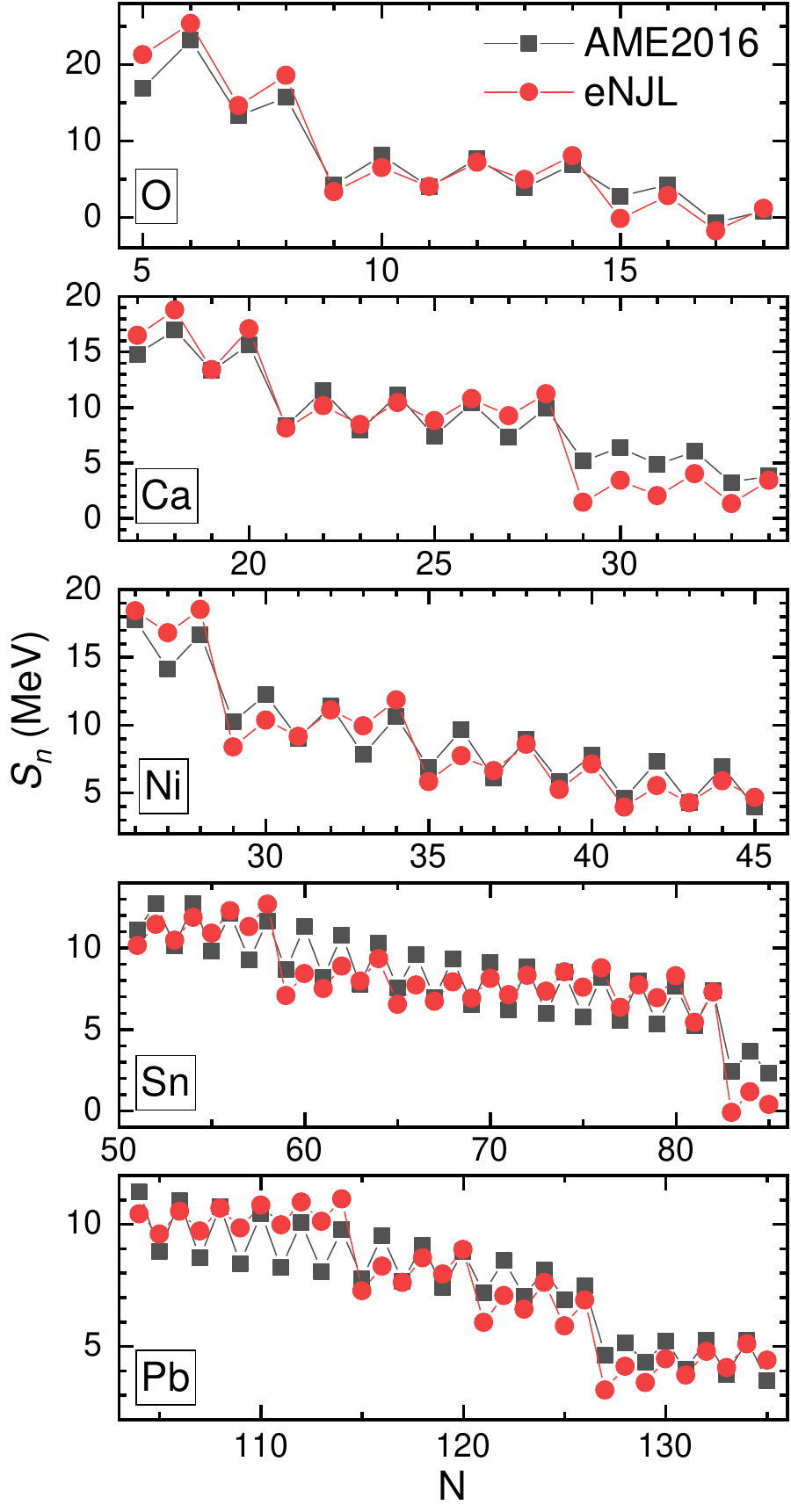}
  \caption{\label{Fig:Sn} Single neutron separation energies $S_n$ in O, Ca, Ni, Sn, and Pb isotopes calculated with the eNJL model, in comparison with the experimental data~\cite{Audi2017_CPC41-030001, Huang2017_CPC41-030002, Wang2017_CPC41-030003}.}
\end{figure}

In Fig.~\ref{Fig:Sn} we present the single neutron separation energies $S_n = B(Z, A) - B(Z, A-1)$ in O, Ca, Ni, Sn, and Pb isotopes. It is found that our eNJL model generally reproduces the experimental neutron separation energies as well as the trends with respect to the variations of $N$. The odd-even staggering is also obtained and generally consistent with the experimental values, where the simple formula in Eq.~(\ref{EQ:pair}) well describes the pairing correlations in finite nuclei. The shell closures at $N= 8$, 20, 28, 82, and 126 can be easily identified with the sudden decrease of $S_n$ as we increase $N$. Nevertheless, the extent of decline in $S_n$ is slightly larger than the experimental values, which is mainly attributed to the too large shell gaps at $N= 28$, 82, and 126. The spurious shell closures at $N= 34$, 58, 92, and 114 are also identified with unrealistic decline in $S_n$, leading to slight deviations of the obtained neutron separation energies from the experimental values.

\begin{figure}[!ht]
  \centering
  \includegraphics[width=\linewidth]{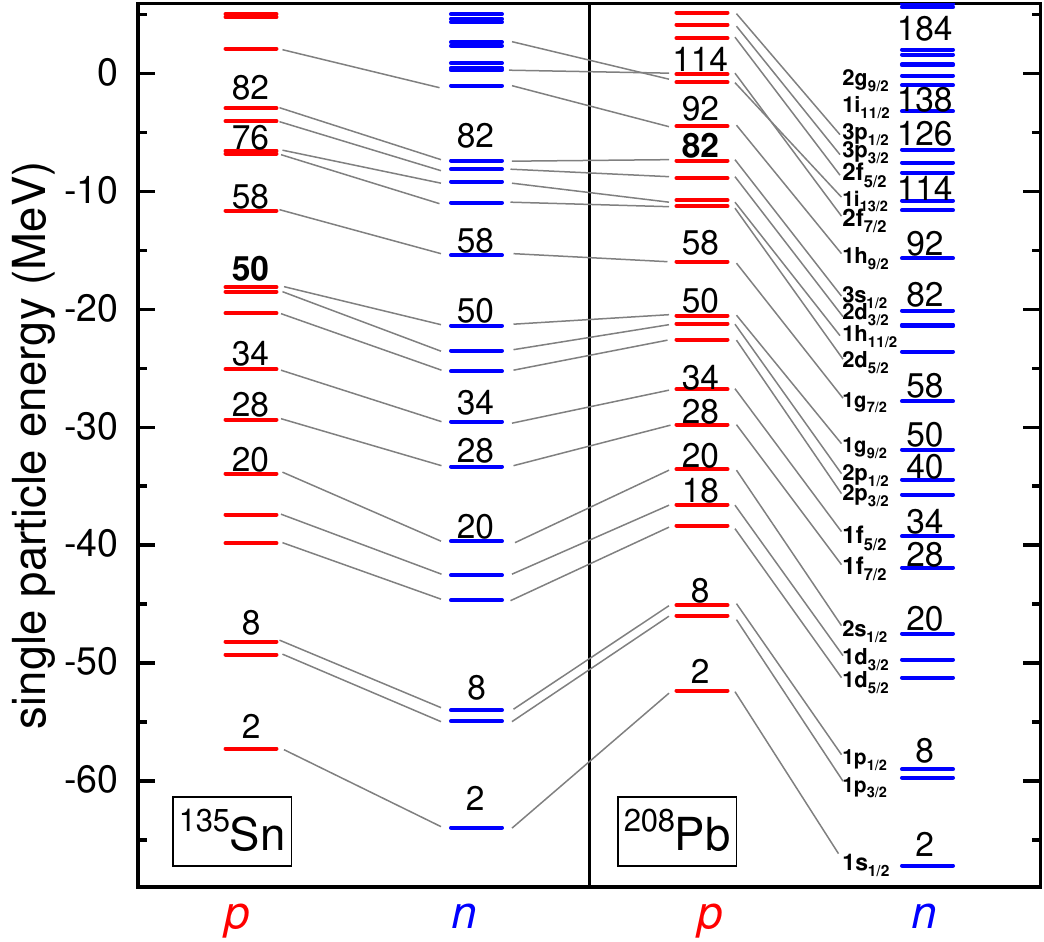}
  \caption{\label{Fig:spl} Single particle energies of protons and neutrons in $^{135}$Sn and $^{208}$Pb predicted by the eNJL model.}
\end{figure}

To show the (spurious) shell closures predicted by our eNJL model more clearly, in Fig.~\ref{Fig:spl} we present the single particle energies of protons and neutrons in $^{135}$Sn and $^{208}$Pb. It is found that our eNJL model generally reproduces the magic numbers in nuclei at $N(Z)=2$, 8, 20, 28, 50, 82, and 126, while spurious shell closures is identified as well at $N(Z)=34$, 58, 92, and 114. Combined with the single neutron separation energies $S_n$ in O isotopes, it can be easily identified that the spin-orbit splitting of the level 1$d$ is too large, leading to a slightly too large decline in $S_n$ at $N=14$. Similar situation is observed for other orbitals, e.g., the spacings in the pairs of levels 1$f_{7/2}$-1$f_{5/2}$, 1$f_{5/2}$-2$p_{3/2}$, 1$g_{7/2}$-2$d_{5/2}$, 3$s_{1/2}$-1$h_{9/2}$, 1$i_{13/2}$-2$f_{5/2}$, 2$f_{5/2}$-3$p_{3/2}$, and 3$p_{1/2}$-1$i_{11/2}$ are too large. This problem may be solved in our future work by carefully calibrating the density dependence of coupling constants so that the too large level spacings can be reduced with the spurious shell closures eliminated~\cite{Wei2020_CPC44-074107}.

\begin{figure}[!ht]
  \centering
  \includegraphics[width=\linewidth]{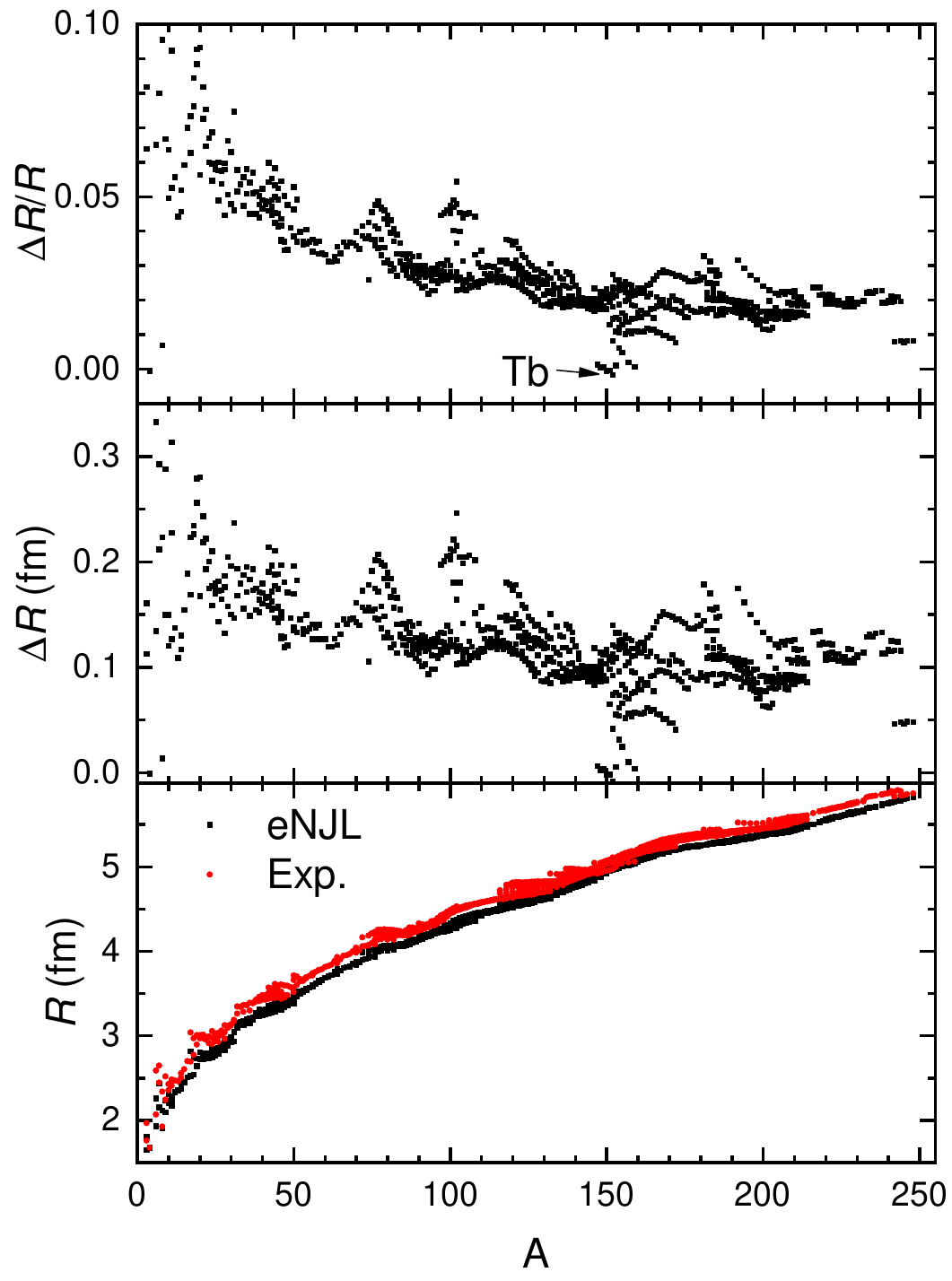}
  \caption{\label{Fig:Drch1D} Charge radii $R$ of 906 nuclei predicted by the eNJL model, in comparison with the experimental data~\cite{Angeli2013_ADNDT99-69}.}
\end{figure}

Finally, in Fig.~\ref{Fig:Drch1D} we present the charge radii $R$ of 906 nuclei predicted by the eNJL model, the differences $\Delta R = R-R_\mathrm{exp}$ and relative differences $\Delta R/R$ with respect to the experimental values $R_\mathrm{exp}$~\cite{Angeli2013_ADNDT99-69} are indicated in the upper panels as well. In general, the charge radii of 906 nuclei predicted by our eNJL model are systematically smaller than the experimental values with the root-mean-square deviations $\sqrt{\sum_i \Delta R_i^2/906} = 0.127$ fm. This could be the reason of too large level spacings as indicated in Fig.~\ref{Fig:spl}, which should be improved in our future investigations.

\section{\label{sec:con}Conclusion}

Assuming spherical symmetry and neglecting the center-of-mass corrections, we investigate systematically the properties of finite nuclei based on an extended NJL (eNJL) model that treats baryons as clusters of quarks~\cite{Xia2024_PRD110-014022, Xia2024}. The Dirac sea, the spontaneous chiral symmetry breaking, and the quark degrees of freedom are considered self-consistently in the eNJL model.
The binding energies of the 2495 nuclei ($A > 2$) from AME2016~\cite{Audi2017_CPC41-030001, Huang2017_CPC41-030002, Wang2017_CPC41-030003} are reproduced using the eNJL model with the root-mean-square deviations 5.38 MeV. It is found that the deviations are  mainly attributed to the too large shell gaps at $N(Z)$ = 28, 50, 82, 126 and 34, 58, 92. The charge radii of 906 nuclei are obtained and compared with the experimental values, which are systematically smaller with the root-mean-square deviations 0.127 fm. In our future works, by carefully calibrating the density dependence of coupling constants and considering deformations with microscopic collective corrections from the nucleons in the Fermi sea and quarks in the Dirac sea, we expect the deviations can be further reduced.

\begin{acknowledgments}
This work was supported by the National Natural Science Foundation of China (Grant Nos. 12275234, 12205057, U2032141), the National SKA Program of China (Grant No. 2020SKA0120300), the Natural Science Foundation of Henan Province (Grant No. 242300421156), and the Science and Technology Plan Project of Guangxi (Grant No. Guike AD23026250).
\end{acknowledgments}


%

\end{document}